\documentclass[twocolumn,showpacs]{revtex4}
\usepackage{epsfig}
\usepackage{graphicx}
\input{psfig.sty}

\newcommand{\bastar}{\begin{eqnarray*}}
\newcommand{\eastar}{\end{eqnarray*}}
\newskip\humongous \humongous=0pt plus 1000pt minus 1000pt

\newif\ifdtup

\relax
\newcommand{\be}{\begin{equation}}
\newcommand{\ee}{\end{equation}}
\newcommand{\bea}{\begin{eqnarray}}
\newcommand{\eea}{\end{eqnarray}}

\newcommand{\dfrac}{\displaystyle\frac}
\newcommand{\ba}{\begin{array}}
\newcommand{\ea}{\end{array}}

\newcommand{\nn}{\nonumber}

\begin{document}
\title{Analytic Electroweak Dyon}
\bigskip

\author{Y. M. Cho}
\email{ymcho@yongmin.snu.ac.kr}
\affiliation{Department of Physics, College of Natural Sciences,
Seoul National University, Seoul 151-742, Korea}
\begin{abstract}
We present analytic monopole and dyon
solutions whose energy is fixed by the electroweak scale.
Our result shows that genuine
electroweak monopole and dyon could exist whose
mass scale is much smaller than the grand unification scale.
\end{abstract}
\maketitle

Ever since Dirac \cite{Dirac} has introduced the concept of the
magnetic monopole, the monopoles have remained a fascinating subject
in theoretical physics.
The Abelian monopole has been generalized to the
point-like non-Abelian monopole by Wu and Yang \cite{Wu,cho80},
and to the finite energy soliton
by 't Hooft and Polyakov \cite{Hooft,Julia}.

In the interesting case of electroweak theory of Weinberg and Salam,
however, it has generally been believed that there
exists no topological monopole of physical interest.
The basis for this ``non-existence theorem'' is, of course, that with
the spontaneous symmetry breaking the quotient space $SU(2) \times
U(1)/U(1)_{\rm em}$ allows no non-trivial second homotopy.
This belief, however, is unfounded. Indeed, recently
Cho and Maison~\cite{Cho0,cho97} have established
that Weinberg-Salam model has exactly the same topological
structure as Georgi-Glashow model,
and demonstrated the existence of a new type of
monopole and dyon solutions in the standard
electroweak theory.
This was based on the observation that
Weinberg-Salam model, with the hypercharge $U(1)$,
could be viewed as a gauged $CP^1$ model in which the (normalized)
Higgs doublet plays the role of the $CP^1$ field. So
Weinberg-Salam model and Georgi-Glashow model
have exactly the same nontrivial second homotopy $\pi_2(S^2)$
which allows topological monopoles.
Originally the solutions of Cho and Maison
were obtained by a numerical integration.
But a mathematically
rigorous existence proof has since been established which endorses
the numerical results, and the solutions are now referred to
as the Cho-Maison monopole and dyon \cite{Yang1,Yang2}.

The Cho-Maison monopole may be viewed as
a hybrid between the Abelian point monopole
and the 't Hooft-Polyakov monopole, because it has a $U(1)$ point
singularity at the center even though the $SU(2)$ part is completely
regular. Consequently it carries an infinite energy
at the classical level, so that
physically the mass of the monopole remains arbitrary.
{\em A priori} there is nothing
wrong with this, but nevertheless one
may wonder whether one can have an analytic
electroweak monopole which has a finite energy.
{\it The purpose of this Letter is to show that this is indeed possible,
and to present finite energy electroweak monopole and dyon solutions
which are analytic everywhere, including the origin}.

Let us start with the Lagrangian of
the standard Weinberg-Salam model,
\begin{eqnarray}
{\cal L} &=& -|{\cal D}_{\mu} \phi |^2
 -\frac{\lambda}{2} (\phi^\dagger \phi
 -\frac{\mu^2}{\lambda} )^2
 -\frac{1}{4} \vec F_{\mu\nu}^2
 -\frac{1}{4} G_{\mu\nu}^2 , \nonumber
\end{eqnarray}
\begin{eqnarray}
{\cal D}_{\mu} \phi &=& \Big(\partial_{\mu}
  -i\frac{g}{2} \vec \tau \cdot \vec A_{\mu}
  -i\frac{g'}{2} B_{\mu}\Big) \phi,
\label{lag1}
\end{eqnarray}
where $\phi$ is the Higgs doublet,
$\vec F_{\mu\nu}$ and $G_{\mu\nu}$
are the gauge field strengths of $SU(2)$ and $U(1)$ with the
potentials $\vec A_{\mu}$ and $B_{\mu}$.
Now we choose the following static spherically symmetric ansatz
\begin{eqnarray}
\mbox{\boldmath$\phi$}&=&\frac{1}{\sqrt{2}}\rho(r)\xi(\theta,\varphi),
  \nonumber \\
\xi&=&i\left(\begin{array}{cc}
         \sin (\theta/2)\,\, e^{-i\varphi}\\
       - \cos(\theta/2)
      \end{array} \right),
\hspace{5mm} \hat n = \xi^{\dagger} \vec \tau ~\xi = - \hat{r},
  \nonumber \\
\vec A_{\mu} &=& \frac{1}{g} A(r) \partial_{\mu}t ~\hat n
   +\frac{1}{g}(f(r)-1) ~\hat n \times
    \partial_{\mu} \hat n,
\label{ansatz1},\\
B_{\mu} &=&-\frac{1}{g'} B(r) \partial_{\mu}t -
            \frac{1}{g'}(1-\cos\theta) \partial_{\mu} \varphi,
\nonumber
\end{eqnarray}
where $(t,r, \theta, \varphi)$ are the spherical coordinates.
Notice that the apparent string
singularity along the negative z-axis in $\xi$ and $B_{\mu}$ is a pure
gauge artifact which can easily be removed with a hypercharge $U(1)$
gauge transformation. So the above ansatz describes a most
general spherically symmetric ansatz of a $SU(2) \times U(1)$ dyon.
Here we emphasize the importance of the non-trivial $U(1)$
degrees to make the ansatz spherically symmetric \cite{Cho0,cho97}.

To understand the physical content of the ansatz we now
perform the following gauge transformation on (\ref{ansatz1})
\begin{eqnarray}
\xi \longrightarrow i\left( \begin{array}{cc}
        \cos (\theta/2)& \sin(\theta/2)e^{-i\varphi} \\
        -\sin(\theta/2) e^{i\varphi} & \cos(\theta/2)
\end{array}
\right) \xi = \left(\begin{array}{cc}
0 \\ 1
\end{array} \right),
\label{gauge}
\end{eqnarray}
and let $\vec A_\mu=(A_\mu^1,~A_\mu^2,~A_\mu^3)$ in this unitary gauge.
Now, introducing the electromagnetic potential $A_\mu^{\rm (em)}$ and the
neutral $Z$-boson $Z_\mu$ with the Weinberg angle $\theta_{\rm w}$
\begin{eqnarray}
&\left( \begin{array}{cc}
A_\mu^{\rm (em)} \\  Z_{\mu}
\end{array} \right)
= \left(\begin{array}{cc}
\cos\theta_{\rm w} & \sin\theta_{\rm w}\\
-\sin\theta_{\rm w} & \cos\theta_{\rm w}
\end{array} \right)
\left( \begin{array}{cc}
B_{\mu} \\ A^3_{\mu}
\end{array} \right) \nonumber\\
&= \dfrac{1}{\sqrt{g^2 + g'^2}}
\left(\begin{array}{cc}
g & g' \\ -g' & g
\end{array} \right)
\left( \begin{array}{cc}
B_{\mu} \\ A^3_{\mu}
\end{array} \right) , \label{wein}
\end{eqnarray}
we can express the ansatz (\ref{ansatz1}) by
\begin{eqnarray}
\label{ansatz2}
\rho &=& \rho(r) \nn\\
W_{\mu} &=& \dfrac{1}{\sqrt{2}}(A_\mu^1 + i A_\mu^2) \nn\\
&=&\frac{i}{g}\frac{f(r)}{\sqrt2}e^{i\varphi}
      (\partial_\mu \theta +i \sin\theta \partial_\mu \varphi), \nn\\
A_{\mu}^{\rm (em)} &=& - e \Big( \frac{A(r)}{g^2} +
    \frac{B(r)}{g'^2} \Big) \partial_{\mu}t
-\frac{1}{e}(1-\cos\theta) \partial_{\mu} \varphi,  \nonumber \\
Z_{\mu} &=& \frac{e}{gg'}(B(r)-A(r)) \partial_{\mu}t,
\end{eqnarray}
where $\rho$ and $W_\mu$ are Higgs boson and $W$-boson,
and $e$ is the electric charge
\begin{eqnarray}
e=\frac{gg'}{\sqrt{g^2+g'^2}}=g\sin\theta_{\rm w}.
\end{eqnarray}
This clearly shows that the ansatz is
for the electromagnetic monopole and dyon.

With the spherically symmetric ansatz
and with a proper boundary condition one can obtain
the Cho-Maison dyon solution shown in Fig.1, which has
the magnetic charge $4\pi/e$~\cite{Cho0}.
The regular part of the solution
looks very much like the Julia-Zee dyon,
except that it  has a non-trivial
$Z$-boson dressing. Of course the magnetic singularity at
the origin makes the energy of the Cho-Maison solutions infinite.
A simple way to make the energy finite is to introduce
the gravitational interaction~\cite{Bais}.
But the gravitational interaction is not likely remove the singularity
at the origin.

To construct the analytic
monopole and dyon solutions,
notice that non-Abelian gauge theory in general
is nothing but a special type of an Abelian gauge theory
which has a well-defined set of charged vector fields as its source.
This tells that the finite energy non-Abelian monopoles
are really nothing but the
Abelian monopoles whose singularity is regularized
by the charged vector fields \cite{cho97,klee}.
From this perspective one can try to
make the energy of the above solutions
finite by introducing additional interactions and/or charged vector fields.

It is rather straightfoward to obtain a finite energy dyon
solution by introducing additional hypercharged vector fields.
This can be done by enlarging the hypercharge $U(1)$ to another
$SU(2)$ and extending the gauge group to $SU(2) \times SU(2)$
\cite{cho97}. But a more economic way to obtain a finite energy electroweak
dyon is utilizing the already existing $W$-boson.
In this case we could try to regularize
the magnetic singularity of the Cho-Maison solutions
with a judicious choice of an extra
electromagnetic interaction of $W$-boson
with the monopole.

To show that this is indeed possible notice
that in the unitary gauge (\ref{gauge}) where $\hat n$ assumes
the trivial configuration $(0,0,-1)$, the Lagrangian (\ref{lag1})
can be written as
\begin{eqnarray}
&{\cal L} =-\dfrac{1}{2}(\partial_\mu \rho)^2
-\dfrac{\lambda}{8}\Big(\rho^2-\dfrac{2\mu^2}{\lambda}\Big)^2
-\dfrac{g^2}{4} \rho^2W_\mu^* W_\mu \nn\\
& -\dfrac{1}{2}|(D_\mu^{\rm (em)} W_\nu - D_\nu^{\rm (em)} W_\mu)
+ie \dfrac{g}{g'}(Z_\mu W_\nu - Z_\nu W_\mu)|^2 \nn\\
&-\dfrac{1}{4} {F_{\mu\nu}^{\rm (em)}}^2
+ ie F_{\mu\nu}^{\rm (em)} W_\mu^* W_\nu
+ \dfrac{g^2}{4}(W_\mu^* W_\nu - W_\nu^* W_\mu)^2 \nn\\
&-\dfrac{1}{4} Z_{\mu\nu}^2
-\dfrac{g^2+g'^2}{8} \rho^2 Z_\mu^2
+\dfrac{ig^2}{\sqrt{g^2+g'^2}} Z_{\mu\nu} W_\mu^* W_\nu,
\label{lag2}
\end{eqnarray}
where $Z_{\mu\nu}=\partial_\mu Z_\nu -\partial_\nu Z_\mu$
and $D_\mu^{\rm (em)} =\partial_\mu +ig A_\mu^{\rm (em)}$.
We now introduce an extra interaction ${\cal L}'$,
\begin{eqnarray}
{\cal L}'=i\alpha gF_{\mu\nu}^{\rm (em)} W_\mu^* W_\nu
+\beta\dfrac{g^2}{4}(W_\mu^*W_\nu-W_\nu^*W_\mu)^2.
\label{int1}
\end{eqnarray}
With this additional interaction the energy of the dyon
is given by $E=E_0 +E_1$, where
\begin{eqnarray}
E_0&=&
\frac{2\pi}{g^2}\int\limits_0^\infty
\frac{dr}{r^2}\bigg\{
\frac{g^2}{g'^2}+1-2(1+\alpha)f^2+(1+\beta)f^4
\bigg\} \nonumber,\\
E_1&=&\frac{4\pi}{g^2} \int\limits_0^\infty dr \bigg\{
(\dot f)^2 + \frac{g^2}{2}(r\dot\rho)^2 + \frac{1}{2}(r\dot A)^2
+ \frac{g^2}{2g'^2}(r\dot B)^2 \nn\\
&+& \frac{g^2}{4} f^2\rho^2 + f^2 A^2
+ \dfrac{g^2r^2}{8} (B-A)^2 \rho^2 \nn\\
&+& \dfrac{\lambda g^2r^2}{8}\Big(\rho^2-\frac{2\mu^2}{\lambda}\Big)^2
\bigg\}.
\end{eqnarray}
Clearly $E_1$ could be made  finite with a proper boundary condition,
but notice that when $\alpha=\beta=0$, $E_0$ becomes infinite.
This is the reason why the Cho-Maison dyon has infinite energy.
To make $E_0$ finite we need to remove both $1/r^2$ and $1/r$
singularities in $E_0$. This requires
\begin{eqnarray}
&1+\dfrac{g^2}{g'^2}-2(1+\alpha) f^2(0)+(1+\beta) f^4(0)=0, \nn\\
&(1+\alpha)f(0)-(1+\beta) f^3(0) =0.
\end{eqnarray}
Thus we must have
\begin{eqnarray}
\label{cond3}
&\dfrac{(1+\alpha)^2}{1+\beta}
=1+\dfrac{g^2}{g'^2}=\dfrac{1}{\sin^2\theta_{\rm w}}, \nonumber\\
&f(0)=\dfrac{1}{\sqrt{(1+\alpha)\sin^2\theta_{\rm w}}}.
\end{eqnarray}
With this we have the following equations of motion
\begin{eqnarray}
&&\ddot f - \dfrac{(1+\alpha)}{r^2}\Big(\frac{f^2}{f^2(0)}-1\Big) f
=\Big(\frac{g^2}{4}\rho^2-A^2 \Big) f, \nonumber \\
&&\ddot \rho+\frac{2}{r}\dot\rho
-\frac{f^2}{2r^2}\rho
=-\frac{1}{4}(B-A)^2 \rho
+\dfrac{\lambda}{2} \Big(\rho^2 -\frac{2\mu^2}{\lambda}\Big)\rho,
 \nonumber \\
&&\ddot A +\frac{2}{r} \dot A
- \frac{2f^2}{r^2} A
=\frac{g^2}{4}(A-B)\rho^2 ,
\label{eqm3} \\
&&\ddot B+\frac{2}{r}\dot B
=\frac{g'^2}{4}(B-A) \rho^2 ,
  \nonumber
\end{eqnarray}
which can be integrated
with the boundary condition
\begin{eqnarray}
&&f(0)=1/\sqrt{(1+\alpha)\sin^2\theta_{\rm w}},~~\rho(0)=0, \nn\\
&&A(0)=0,~~B(0)=b_0, \nn\\
&&f(\infty)=0,~~\rho(\infty)=\rho_0 =\sqrt{2\mu^2/\lambda}, \nn\\
&&A(\infty)=B(\infty)=A_0.
\label{bc1}
\end{eqnarray}
But notice that, although obviously sufficient for a finite energy
solution, the condition (\ref{bc1}) in general does not guarantee the
analyticity of the gauge potential at the origin. This must be clear
from the fact that the condition (\ref{bc1}) does not remove
the singularity in $B_\mu$ \cite{cho97}.

The condition for an analytic solution is given by
\bea
\label{bc2}
\alpha = 0, ~~~~~f(0)=\frac{1}{\sin \theta_{\rm w}}=\dfrac{g}{e}.
\eea
To understand this we need to review the analytic
Julia-Zee dyon in Georgi-Glashow model
\begin{eqnarray}
\label{eq:Julia}
{\cal L}_{GG} =-\dfrac{1}{2}(D_\mu \vec \Phi)^2
-\dfrac{\lambda}{4}\left(\vec \Phi^2-\frac{\mu^2}{\lambda}\right)^2
-\dfrac{1}{4} \vec F_{\mu\nu}^2 ,
\end{eqnarray}
where $\vec \Phi$ is the Higgs triplet.
With $\vec \Phi = \rho \hat n$
one can easily show that the Georgi-Glashow model
acquires the following Abelian form in the unitary gauge
\begin{eqnarray}
\label{eq:lagran}
&{\cal L}_{GG} = -\dfrac{1}{2} (\partial_\mu \rho)^2
-\dfrac{\lambda}{4}\left(\rho^2 -\frac{\mu^2}{\lambda}\right)^2
- g^2 {\rho}^2 W_\mu^*W_\mu \nn\\
&- \dfrac{1}{4} F_{\mu\nu}^2
+ ig F_{\mu\nu} W_\mu^*W_\nu
+ \dfrac{g^2}{4}(W_\mu^* W_\nu - W_\nu^* W_\mu)^2 \nn\\
&-\dfrac{1}{2} |D_\mu W_\nu - D_\nu W_\mu|^2.
\end{eqnarray}
Now, with the spherically symmetric ansatz
\begin{eqnarray}
\label{eq:Zee}
\Phi&=&\rho(r)\hat{n} \nonumber,\\
\vec A_\mu&=&\frac{1}{g}A(r)\partial_\mu t\,\,\hat{n}
+ \frac{1}{g}(f(r)-1)\hat{n}\times\partial_\mu \hat{n},
\end{eqnarray}
one has the following equation
\begin{eqnarray}
\label{eq:JuliaZee}
&&\ddot{f}- \frac{f^2-1}{r^2}f=\left(g^2\rho^2-A^2\right)f, \nonumber\\
&&\ddot{\rho}+\frac{2}{r}\dot{\rho} - \frac{2f^2}{r^2}\rho =
     \lambda \Big(\rho^2 - \frac{\mu^2}{\lambda} \Big)\rho , \\
&&\ddot{A}+\frac{2}{r}\dot{A} -\frac{2f^2}{r^2} A=0. \nonumber
\end{eqnarray}
With the boundary condition
\bea
&f(0)=1,~~~~~\rho(0)=0,~~~~~A(0)=0, \nn\\
&f(\infty)=0,~~~~~\rho(\infty)=\rho_0,~~~~~A(\infty)=A_0,
\eea
one can integrate (\ref{eq:JuliaZee}) and obtain
the Julia-Zee dyon. Notice that the boundary condition,
in particular $f(0)=1$, is crucial to make the ansatz (17)
analytic at the origin.

\begin{figure}
\epsfysize=5.5cm \centerline{\epsffile{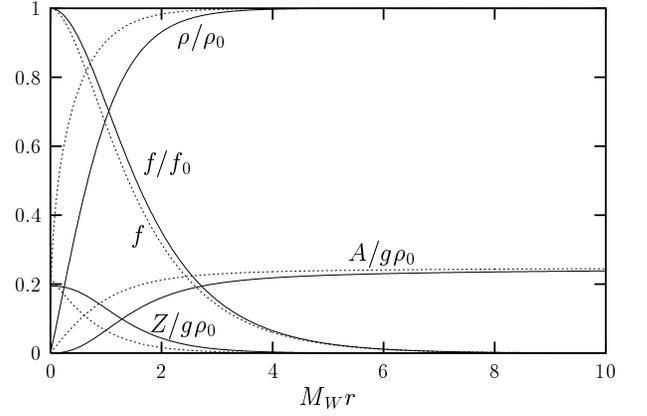}} \caption{The
electroweak dyon solutions. The solid line represents the finite
energy dyon and dotted line represents the Cho-Maison dyon, where
$Z=B-A$ and we have chosen $\sin^2\theta_{\rm w}=0.2325$,
$\lambda/g^2=M_H^2/4M_W^2=1/2$, $A(\infty)=M_W/2$.} \label{fig1}
\end{figure}

To derive the analyticity condition (\ref{bc2})
notice that, with (\ref{bc2}), what
the extra interaction (\ref{int1}) does is to modify
the coupling strength of the $W$-boson quartic self-interaction
from $g^2/4$ to $e^2/4$.
So, in the absence of the $Z$-boson we have
\begin{eqnarray}
\label{Lag0}
&{\cal L}+{\cal L'}
\rightarrow -\dfrac{1}{2}(\partial_\mu \rho)^2
-\dfrac{\lambda}{8}\Big(\rho^2-\dfrac{2\mu^2}{\lambda}\Big)^2
- \dfrac{g^2}{4} \rho^2W_\mu^* W_\mu \nn\\
&-\dfrac{1}{4} {F_{\mu\nu}^{\rm (em)}}^2
+ ie F_{\mu\nu}^{\rm (em)} W_\mu^* W_\nu
+ \dfrac{e^2}{4}(W_\mu^* W_\nu - W_\nu^* W_\mu)^2 \nn\\
&-\dfrac{1}{2}|D_\mu^{\rm (em)} W_\nu - D_\nu^{\rm (em)} W_\mu|^2 \nn\\
&= -\dfrac{1}{2}(\partial_\mu \rho)^2
-\dfrac{\lambda}{8}\Big(\rho^2-\dfrac{2\mu^2}{\lambda}\Big)^2
- \dfrac{g^2}{4} \rho^2W_\mu^* W_\mu \nn\\
&-\dfrac{1}{4} \mbox{\boldmath $F$}_{\mu\nu}^2,
\end{eqnarray}
where now $\mbox{\boldmath $F$}_{\mu\nu}$ is the ``electromagnetic''
$SU(2)$ gauge field made of $W_\mu^1$,~$W_\mu^2$,
and $A_\mu^{\rm (em)}$, with the gauge coupling constant $e$.
Furthermore, with $Z_\mu=0$, the ansatz (\ref{ansatz1}) is written as
\bea
\label{ansatz0}
\rho&=&\rho(r), \nn\\
\mbox{\boldmath $A$}_\mu &=& \dfrac{e}{g^2+g'^2} A(r)
~\partial_{\mu}t ~\hat n \nn\\
&+& \dfrac{1}{e}(\dfrac{e}{g}f(r)-1) ~\hat{n}
\times \partial_{\mu} \hat{n}.
\eea
Evidently (\ref{Lag0}) and (\ref{ansatz0}) are
almost identical to (\ref{eq:lagran}) and (\ref{eq:Zee})
of Georgi-Glashow model. In particular, the Yang-Mills part
is completely identical, except that here the
coupling constant is $e$, not $g$. This means that
$\mbox{\boldmath $A$}_\mu$ becomes smooth at the origin
when $A(0)=0$ and $f(0)=g/e$. Furthermore, since $Z_\mu$ has no
monopole singularity, the ansatz (\ref{ansatz1})
becomes smooth everywhere when in this case.
This provides the analyticity condition (\ref{bc2}).

With (\ref{bc1}) and (\ref{bc2}) we can integrate (\ref{eqm3}).
The results of the numerical integration for the dyon solution
are shown in Fig.\ref{fig1}.
Here we have chosen $\sin^2\theta_{\rm w}$ to be the experimental value
$0.2325$.
{\em It is really remarkable that
the finite energy solutions look
almost identical to the Cho-Maison solutions,
even though they no longer have the singularity at the origin
and analytic everywhere. The reason for this must be clear.
All that we need to make the Cho-Maison solutions
analytic is a simple modification of the coupling strength of
$W$-boson quartic self-interaction
from $g^2/4$ to $e^2/4$}.

Clearly the energy of the above solutions
must be of the order of the electroweak scale $M_W=g\rho_0/2$.
Indeed for the monopole the energy with
$\lambda/g^2=1/2$ is given by
\begin{eqnarray}
E=1.407\times\dfrac{4\pi}{e^2}M_W.
\end{eqnarray}
This demonstrates that the finite energy solutions
are really nothing but the regularized
Cho-Maison solutions which have a mass of the electroweak scale.

Notice that we can even find
an analytic monopole solution explicitly,
if we add an extra term $\delta {\cal L}$
to ${\cal L}+{\cal L'}$,
\begin{eqnarray}
\label{mass}
\delta{\cal L}=-\Big(e^2 -\frac{g^2}{4}\Big) \rho^2W_\mu^* W_\mu.
\label{int2}
\end{eqnarray}
This amounts to changing the mass of $W$-boson
from $g\rho_0/2$ to $e \rho_0$. With this change
the electroweak Lagrangian, in the absence of the $Z$-boson,
becomes identical to (\ref{eq:lagran}) in the limit $\lambda=0$.
In this case we have the Bogomol'nyi-Prasad-Sommerfield
equation for the monopole (with $Z_\mu=0$),
\begin{eqnarray}
&&\dot{f} + e \rho f=0,
\nonumber \\
&&\dot{\rho} + \dfrac{1}{er^2} \Big(\frac{f^2}{f(0)^2} -1 \Big)=0.
\label{self2}
\end{eqnarray}
This has the well-known analytic solution \cite{Julia}
\begin{eqnarray}
&f = f(0)\dfrac{e\rho_0 r}{\sinh(e\rho_0r)}
= \dfrac{g\rho_0 r}{\sinh(e\rho_0r)}, \nn\\
&\rho= \rho_0\coth(e\rho_0r)-\dfrac{1}{er},
\end{eqnarray}
which has the energy
$(4\pi/e^2) M'_{W},~(M'_{W}=e\rho_0)$.
But notice that, even in this case,
the electroweak dyon becomes different from Prasad-Sommerfield
dyon, because of the non-trivial $Z$-boson dressing.

Strictly speaking the finite energy solutions are not the solutions of
Weinberg-Salam model, because their existence requires
a modification of the electroweak interaction.
But from the physical point of view there is no doubt that they should
be interpreted as the electroweak monopole and dyon, because
they are really nothing but the regularized Cho-Maison solutions.
More significantly, this regularization is made possible with
only a minor change of the coupling strength of $W$-boson
quartic self-interaction.
From this point of view one could say that, in retrospect,
the existence of the finite energy electroweak dyon
explains why the singular Cho-Maison dyon in Weinberg-Salam model
could exist in the first place.

It has generally been assumed that the finite
energy monopoles could exist only at the grand unification
scale~\cite{Dokos}. {\em But our result suggests the existence of
a totally new type of electroweak
monopole and dyon whose mass is much smaller than the monopoles of the
grand unification}.
Certainly the existence of the finite energy
electroweak monopole and dyon could have important
physical implications.
If existed, they could be the only finite energy topological objects
that one could ever hope to produce with the (future) accelerators.
A more detailed discussion of our work will be published in a separate
paper~\cite{cho02}.

\noindent{\bf Acknowledgments}

The work is supported in part by Korea Research Foundation (Grant KRF-2001
-015-BP0085) and by the BK21 project of Ministry of Education.

\end{document}